\documentclass[reprint,amsmath,amssymb,superscriptaddress,prb]{revtex4-2}

\usepackage{graphicx}
\usepackage{dcolumn}
\usepackage{bm}
\usepackage{color}
\usepackage{amsmath}
\usepackage{booktabs} 
\usepackage{array} 

\begin{document}

\title{Time-dependent density-functional study of hydrogen adsorption and scattering on graphene surfaces}

\author{Samuel S. Taylor}
\affiliation{Department of Physics and Astronomy, Vanderbilt University, Nashville, Tennessee, 37235, USA}

\author{Nicholas Skoufis}
\affiliation{Department of Physics and Astronomy, Vanderbilt University, Nashville, Tennessee, 37235, USA}

\author{Hongbo Du}
\affiliation{Department of Physics, School of Science, Xi’an Technological University, Xi’an, Shaanxi 710032, China}

\author{Cody Covington}
\affiliation{Department of Chemistry, Austin Peay State University,
Clarksville, USA}

\author{K\'alm\'an Varga}
\email{kalman.varga@vanderbilt.edu}
\affiliation{Department of Physics and Astronomy, Vanderbilt University, Nashville, Tennessee, 37235, USA}

\begin{abstract}
Time-dependent density-functional theory simulations are performed to examine the effects of varying incident points and kinetic energies of hydrogen atom projectiles on a graphene-like structure. The simulations reveal that the incident point significantly influences the hydrogen atom's kinetic energy post-interaction, the vibrational dynamics of the graphene lattice, and the scattering angles. Incident points that do not directly collide with carbon atoms result in prolonged interaction times and reduced energy transfer, increasing the likelihood of overcoming the graphene’s potential energy barrier and hydrogen atom adsorption. The study also explores the role of initial kinetic energy in determining adsorption, scattering, or transmission outcomes. These results emphasize the critical influence of initial parameters on the hydrogenation process and provide a foundation for future experimental validation and further exploration of hydrogen-graphene interactions.  

\end{abstract}

\maketitle

\section{Introduction}
Graphene, a material consisting of C atoms arranged in a honeycomb lattice, is well known for its exceptional mechanical strength~\cite{PAPAGEORGIOU201775,doi:10.1126/science.1157996} and remarkable electrical conductivity~\cite{PhysRevB.78.085416,PhysRevB.78.085415}. Graphene's distinctive chemical, electronic, and magnetic properties have made it a focal point of recent research due to its potential applications in electronics~\cite{Novoselov2012,Avouris2012,nano9030374}, solar cells~\cite{MAHMOUDI201851,Das_Sudhagar_Kang_Choi_2014}, and energy storage~\cite{OLABI2021110026,https://doi.org/10.1002/smll.201303202}.

The unique properties of graphene have also driven studies into its chemical modification to induce new functionalities~\cite{C2JM31218B,Ha2012,KUILA20121061,C3NR33708A,C7CS00871F}. One of the most promising modifications is hydrogenation, where hydrogen (H) atoms bond with graphene’s carbon (C) atoms to induce a bandgap, granting graphene the qualities required for semiconductor applications~\cite{Balog2010}. Understanding the interaction of hydrogen with graphene provides critical insights into aspects of graphene’s behavior and material properties, such as hydrogen adsorption mechanisms and the formation of C–H bonds~\cite{doi:10.1126/science.aaw6378,doi:10.1126/science.aax1980,Bonfanti_2018,10.1063/1.3072333,Ivanovskaya2010,JELOAICA1999157,SHA2002318,10.1063/1.1469600,Tachikawa_2010}, vibrational redistribution reactions~\cite{doi:10.1126/science.aax1980,PhysRevB.79.115429}, hydrogen storage~\cite{Jain2020,C2CP42538F,SHIRAZ2017104}, and magnetism~\cite{doi:10.1126/science.aad8038,PhysRevB.77.035427}. Additionally, it provides insights into molecular dynamics~\cite{shi2024quantumdynamicsstudyh,10.1063/5.0176655} and helps identify the conditions that maximize hydrogen adsorption probabilities while minimizing scattering~\cite{doi:10.1126/science.aaw6378,misc/32592884,saito2010incident}

The outcomes of hydrogen–graphene interactions\textemdash adsorption, scattering, or transmission\textemdash depend on several parameters. Key factors include the kinetic energy of the incoming H atom, its incident angle relative to the graphene surface, and the specific incident point on the graphene lattice~\cite{doi:10.1126/science.aaw6378,misc/32592884,saito2010incident}. These parameters critically influence the scattering angles and energy transfer during the interaction, determining the overall dynamics and reaction pathways. To identify the optimal combination of parameters that maximize hydrogen adsorption on the graphene surface, various theoretical and computational methods have been employed\textemdash including classical molecular dynamics (MD) simulations and \textit{ab initio} density functional theory (DFT) simulations.

Classical MD simulations are widely used due to their computational efficiency and ability to sample a large number of conditions~\cite{doi:10.1080/08927020601078471,ITO_NAKAMURA_2006,doi:10.1143/JPSJ.77.114602,10.1063/1.4945034,10.1063/1.4931117,PhysRevB.79.115429,https://doi.org/10.1002/ctpp.200810046,10.1063/1.4813919,misc/32592884,saito2010incident}. For example, classical MD with a reactive empirical bond-order (REBO) potential has been used to explore how varying incident angles and points affect outcomes~\cite{saito2010incident}. However, these methods inherently neglect quantum effects, which have been shown to significantly influence hydrogen–graphene interactions, including altering sticking probabilities~\cite{10.1063/5.0176655}.

DFT is often used as a more precise computational technique for studying H atom and graphene interactions, having been proven accurate in describing adsorption~\cite{JELOAICA1999157,SHA2002318,10.1063/1.1469600,Tachikawa_2010,SHIRAZ2017104} and vibrational relaxation~\cite{PhysRevB.79.115429}. Despite these successes, previous DFT studies have not systematically investigated the role of varying the incident point within graphene’s benzene rings or its impact on scattering angular distributions and energy transfer dynamics.

This study aims to address this gap by employing time-dependent density-functional theory (TDDFT) to investigate the scattering distributions and energy transfer dynamics of hydrogen interacting with graphene. The use of such simulations allows for detailed snapshots and kinetic energy profiles to elucidate the mechanisms underlying hydrogen scattering and C–H bond formation processes. Furthermore, additional simulations explore the influence of varying initial hydrogen velocities while keeping the incident point and angle fixed, providing insight into how kinetic energy affects H atom adsorption, scattering, and transmission reactions.

\section{Computational Method}
The simulations were performed using TDDFT for modeling the electron dynamics on a 
real-space grid with real-time propagation \cite{Varga_Driscoll_2011a}, 
with the Kohn-Sham (KS) Hamiltonian of the following form
\begin{equation}
\begin{split}
\hat{H}_{\text{KS}}(t) = -\frac{\hbar^2}{2m} \nabla^2 + V_{\text{ion}}(\mathbf{r},t) + 
V_{\text{H}}[\rho](\mathbf{r},t) \\
+ V_{\text{XC}}[\rho](\mathbf{r},t).
\end{split}
\label{eq:hamiltonian}
\end{equation}
Here, $\rho$ is the electron density, defined as the sum of the densities of all occupied orbitals:
\begin{equation}
\rho(\mathbf{r},t) = \sum_{k=1}^{\infty} f_k |\psi_k(\mathbf{r},t)|^2,
\end{equation}
where $f_k$ is the occupation number of the orbital $\psi_k$, which can take values 0, 1, or 2 ($f_k = 2$ is allowed due to spin degeneracy). Additionally, $f_k$ must satisfy the constraint $\smash{\sum_{k=1}^{\infty}} f_k = N$, where $N$ is the total number of valence electrons in the system.

$V_{ion}$ in eq.~\ref{eq:hamiltonian} is the external potential due to the ions, represented by employing norm-conserving pseudopotentials centered at each ion as given by Troullier and Martins~\cite{PhysRevB.43.1993}. $V_{H}$ is the Hartree potential, defined as
\begin{equation}
V_H(\mathbf{r}, t) = \int \frac{\rho(\mathbf{r}', t)}{|\mathbf{r} - \mathbf{r}'|} \, d\mathbf{r}',
\end{equation}
and accounts for the electrostatic Coulomb interactions between electrons. The last term in eq.~\ref{eq:hamiltonian}, $V_{XC}$, is the exchange-correlation potential, which is approximated by the adiabatic local-density approximation (ALDA), obtained from a parameterization to a homogeneous electron gas by Perdew and Zunger~\cite{PhysRevB.23.5048}. 

At the beginning of the TDDFT calculations, the ground state of the system is prepared by performing a Density-Functional Theory (DFT) calculation. With these initial conditions in place, we then proceed to propagate the Kohn–Sham orbitals, $\psi_{k}(\mathbf{r},t)$ over time by using the time-dependent KS equation, given as 
\begin{equation}
i \frac{\partial \psi_k(\mathbf{r}, t)}{\partial t} = \hat{H} \psi_k(\mathbf{r}, t).
\label{eq:tdks}
\end{equation}
Eq.~\ref{eq:tdks} was solved using the following time propagator
\begin{equation}
\psi_k(\mathbf{r}, t + \delta t) = \exp\left(-\frac{i \hat{H}_{\text{KS}}(t) \delta t}{\hbar}\right) \psi_k(\mathbf{r}, t).
\end{equation}
This operator is approximated using a fourth-degree Taylor expansion, given as
\begin{equation}
\psi_k(\mathbf{r}, t + \delta t) \approx \sum_{n=0}^{4} \frac{1}{n!} \left(\frac{-i \delta t}{\hbar} \hat{H}_{\text{KS}}(t)\right)^n \psi_k(\mathbf{r}, t).
\end{equation}
The operator is applied for $N$ time steps until the final time, $t_{final} = N \cdot \delta t$, is obtained. A time step of $\delta t = 1$~as was used in the simulations.

In real-space TDDFT, the Kohn-Sham orbitals are represented at discrete points in real space. These points are organized on a uniform rectangular grid. The accuracy of the simulations is determined by the grid spacing, which is the key parameter that can be adjusted. In our simulations, we used a grid spacing of 0.25 \AA\ and placed 81 points along each of the x- and y-axes, and 65 points are placed along the z-axis.

Motion of the ions in the simulations were treated classically. Using the Ehrenfest theorem
, the quantum forces on the ions due to the electrons are given by the derivatives 
of the expectation value of the total electronic energy with respect to the ionic positions. 
These forces are then fed into Newton’s Second Law, giving
\begin{equation}
\begin{split}
M_i \frac{d^2 \mathbf{R}_i}{dt^2} = Z_i \mathbf{E}_{\text{laser}}(t) 
+ \sum_{j \neq i}^{N_{\text{ions}}} \frac{Z_i Z_j (\mathbf{R}_i - \mathbf{R}_j)}{|\mathbf{R}_i - \mathbf{R}_j|^3}\\ 
- \nabla_{\mathbf{R}_i} \int V_{\text{ion}}(\mathbf{r}, \mathbf{R}_i) \rho(\mathbf{r}, t) \, d\mathbf{r},
\end{split}
\end{equation}
where $M_{i}$, $Z_{i}$, and $\mathbf{R}_{i}$ are the mass, pseudocharge (valence), and position of the $i$-th ion, respectively, and $N_{\text{ions}}$ is the total number of ions. This differential equation was time propagated using the Verlet algorithm at every time step $\delta t$.

To model a graphene sheet, coronene (C\textsubscript{24}H\textsubscript{12}) was selected due to its structural similarity to graphene. Coronene, depicted in Fig.~\ref{incident-point-diagram}, consists of 7 benzene rings at its core, resembling the honeycomb lattice of graphene. Its relatively small size allows it to fit comfortably within the simulation grid. 
A larger molecule would significantly increase the grid dimensions and
highly increase computational costs. Graphene super cell with periodic
boundary conditions also computationally expensive because a large
supercell has to be used to avoid the interaction between neighboring
cells and allow scattering with an angle with respect to the surface.

The coronene molecule was positioned at z~=~-2~\AA\ in the xy-plane, and H atom was positioned 7 \AA\ above the coronene at z~=~5~\AA. These positions fit well within the 20~\AA\ $\times$ 20~\AA\ $\times$ 16~\AA\ grid centered at the origin, with sufficient space along the z-axis to accommodate the trajectory of the hydrogen projectile.

In the following section, we present the results of 28 simulations in which a hydrogen projectile was directed at a $4 \times 7$ grid of incident points on the graphene-like target surface. To further investigate the interaction dynamics and explore conditions favorable for hydrogen adsorption, an additional set of 7 simulations was performed with varying projectile kinetic energies at a distinct incident point and angle separate from the first set of 28 simulations.

\section{Results}
\subsection{Incident Points and Scattering}

\begin{figure*}
\centering
\includegraphics[width=\textwidth]{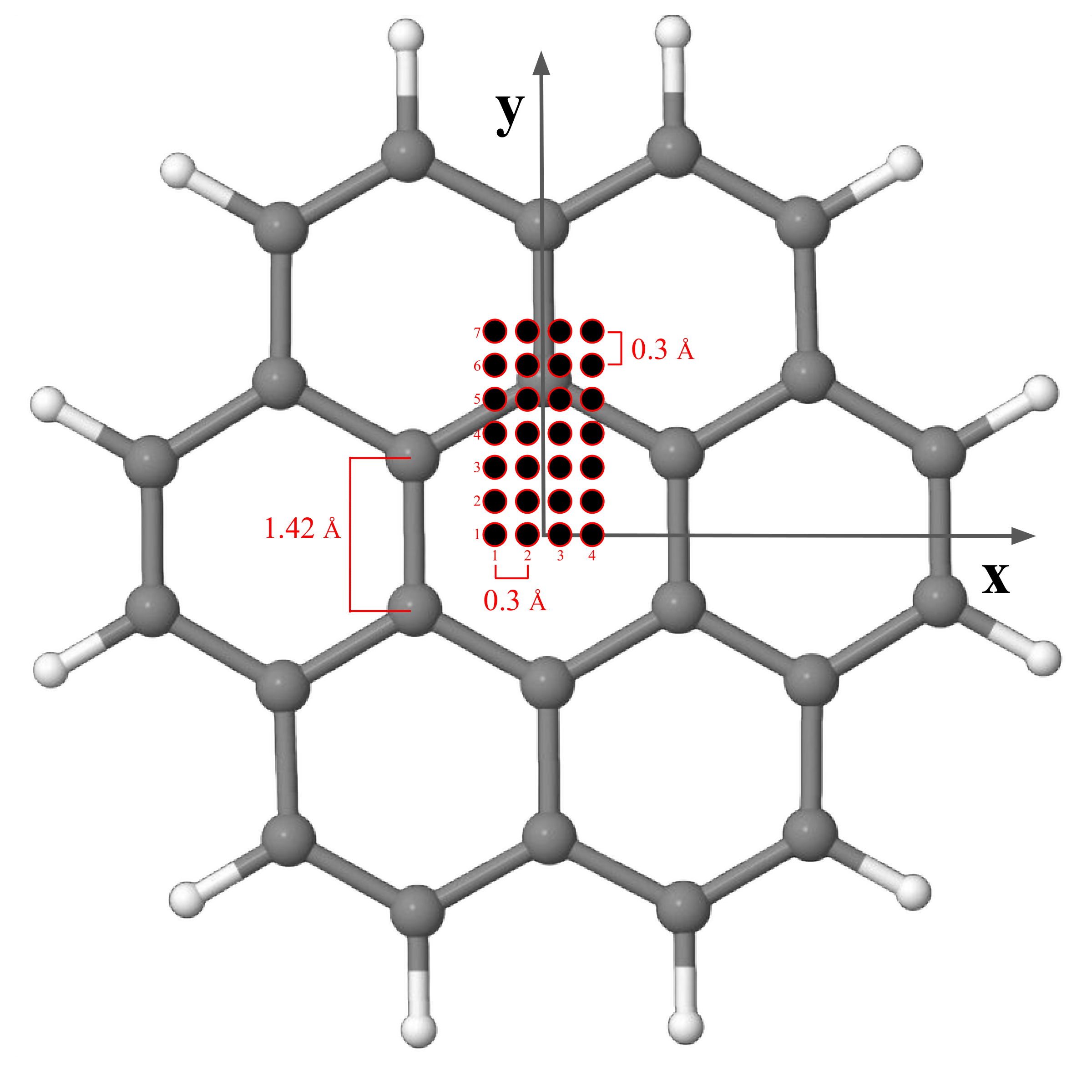}
\caption{Diagram illustrating the incident points of H atom projectiles on the xy-plane of the coronene molecule.}
\label{incident-point-diagram}
\end{figure*}

\begin{figure*}
\centering
\includegraphics[width=\textwidth]{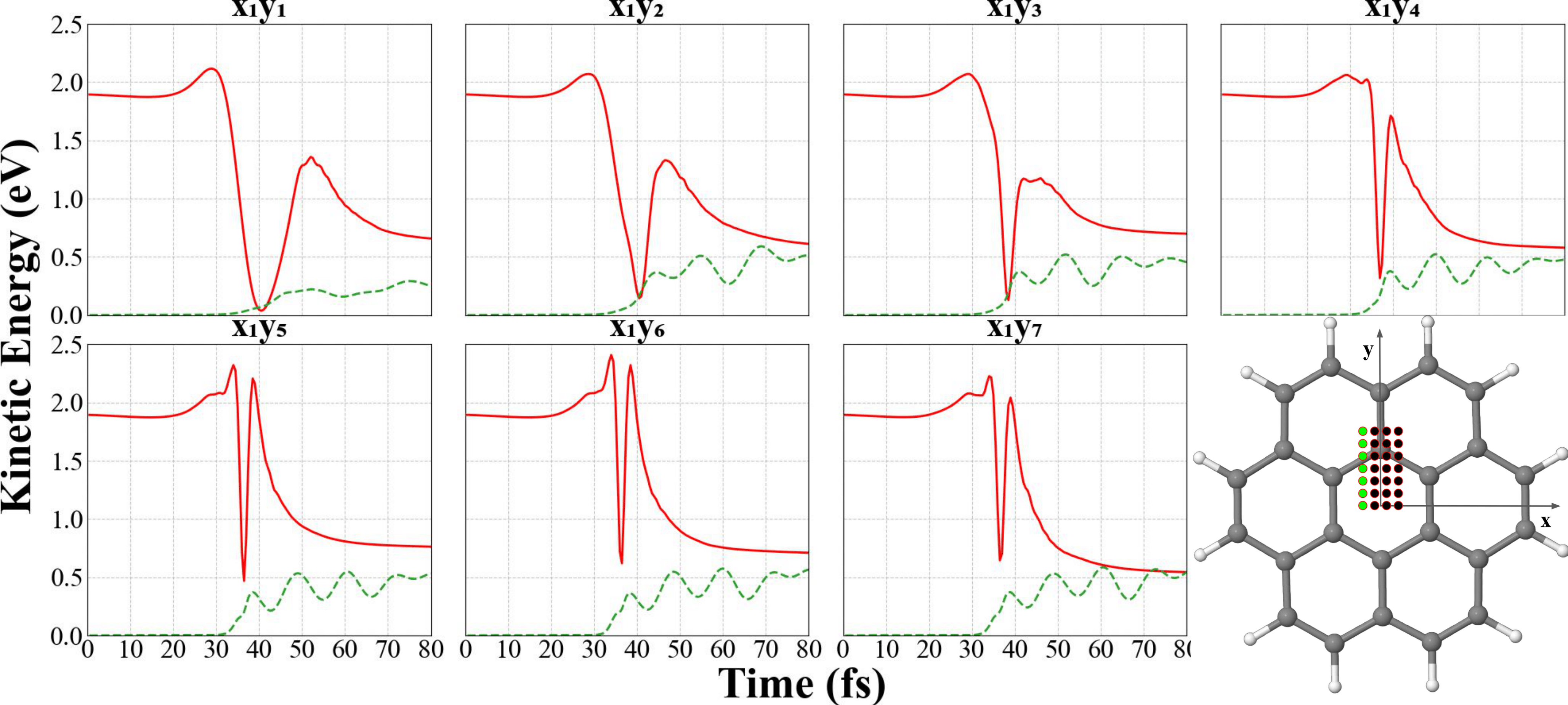}
\caption{Kinetic energy (in eV) of the H atom (red solid line) and the net kinetic energy of all atoms in the coronene sheet (green dashed line) as a function of time throughout the simulation. Each graph corresponds to a different simulation from the column of x\textsubscript{1} incident points, ranging from y\textsubscript{1} to y\textsubscript{7}.}
\label{k-energy-x1}
\end{figure*}

\begin{figure*}
\centering
\includegraphics[width=\textwidth]{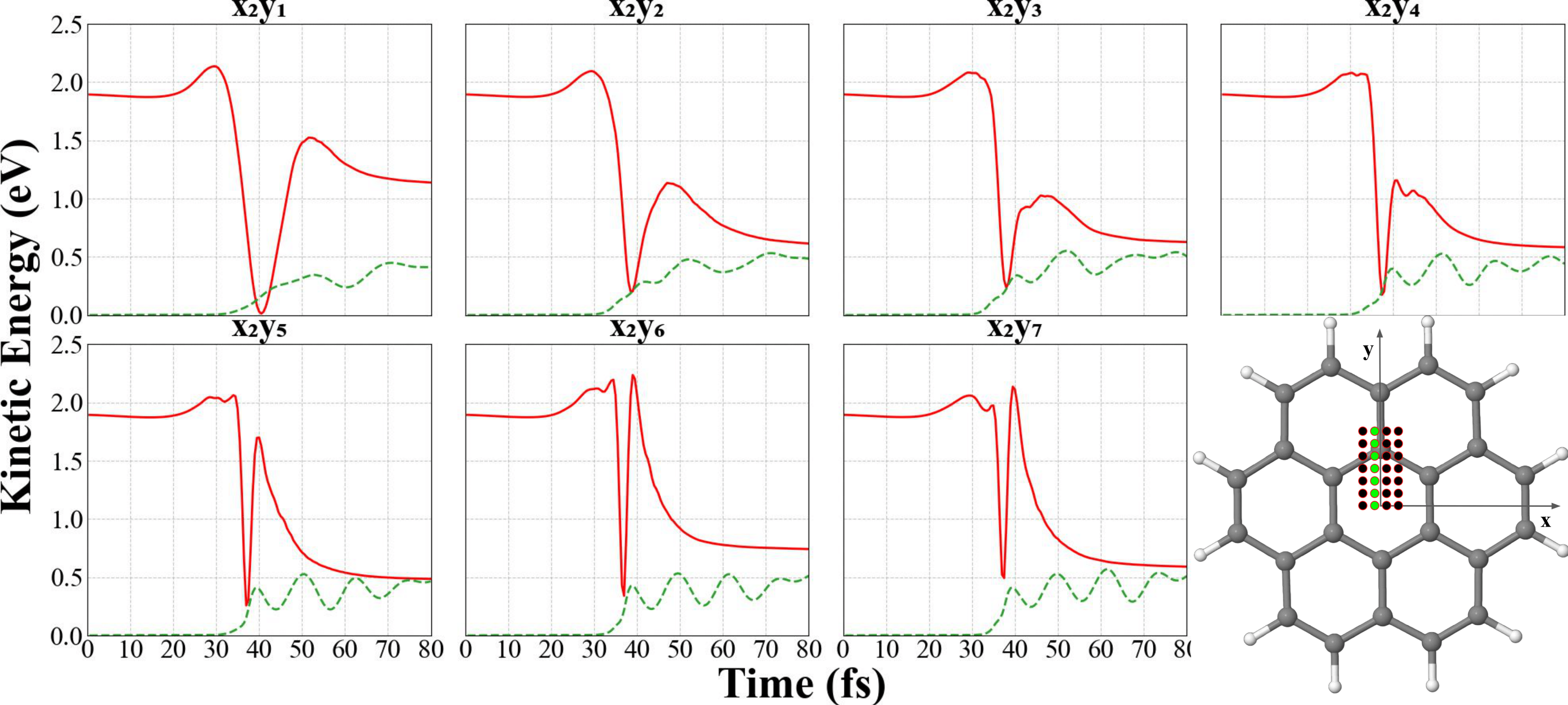}
\caption{Kinetic energy (in eV) of the H atom (red solid line) and the net kinetic energy of all atoms in the coronene sheet (green dashed line) as a function of time throughout the simulation. Each graph corresponds to a different simulation from the column of x\textsubscript{2} incident points, ranging from y\textsubscript{1} to y\textsubscript{7}.}
\label{k-energy-x2}
\end{figure*}

\begin{figure*}
\centering
\includegraphics[width=\textwidth]{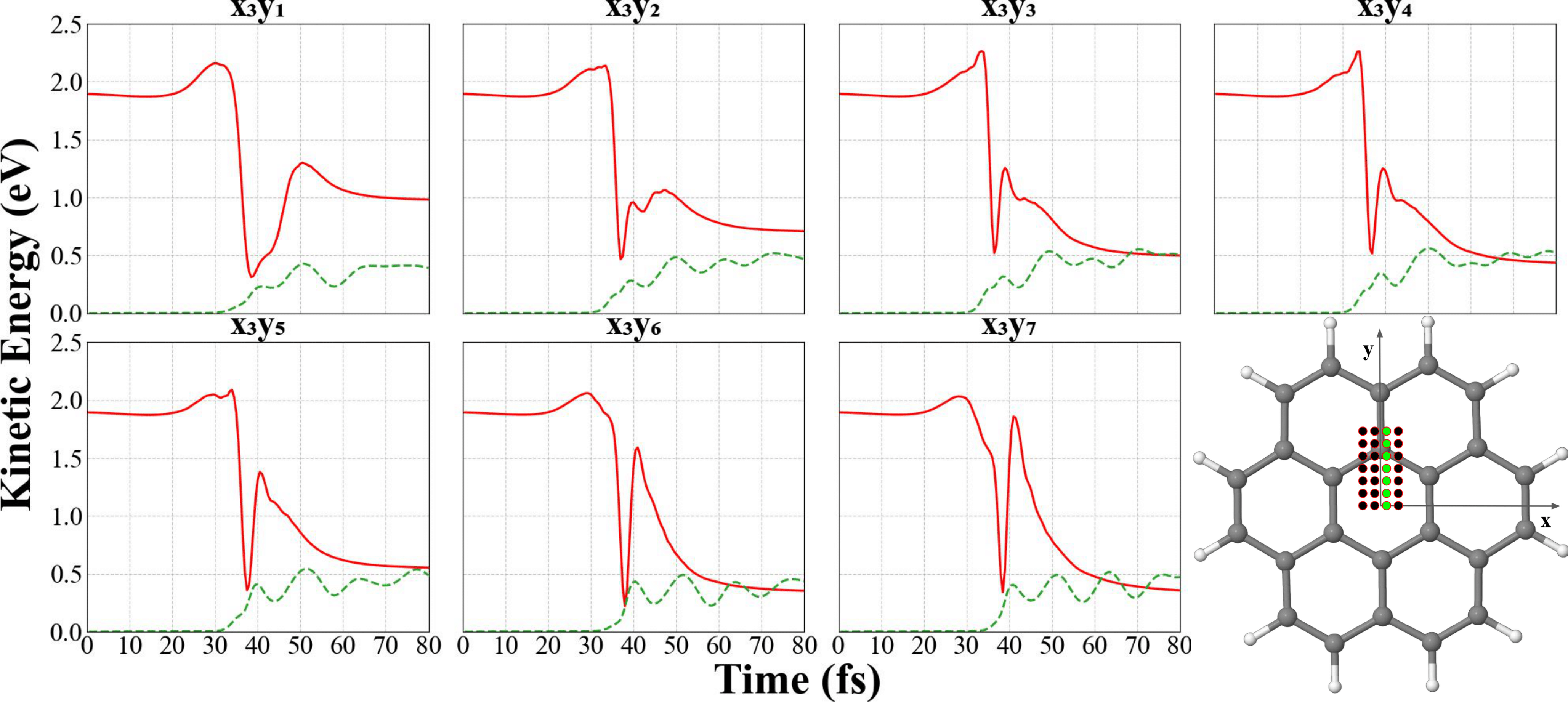}
\caption{Kinetic energy (in eV) of the H atom (red solid line) and the net kinetic energy of all atoms in the coronene sheet (green dashed line) as a function of time throughout the simulation. Each graph corresponds to a different simulation from the column of x\textsubscript{3} incident points, ranging from y\textsubscript{1} to y\textsubscript{7}.}\label{k-energy-x3}
\end{figure*}

\begin{figure*}
\centering
\includegraphics[width=\textwidth]{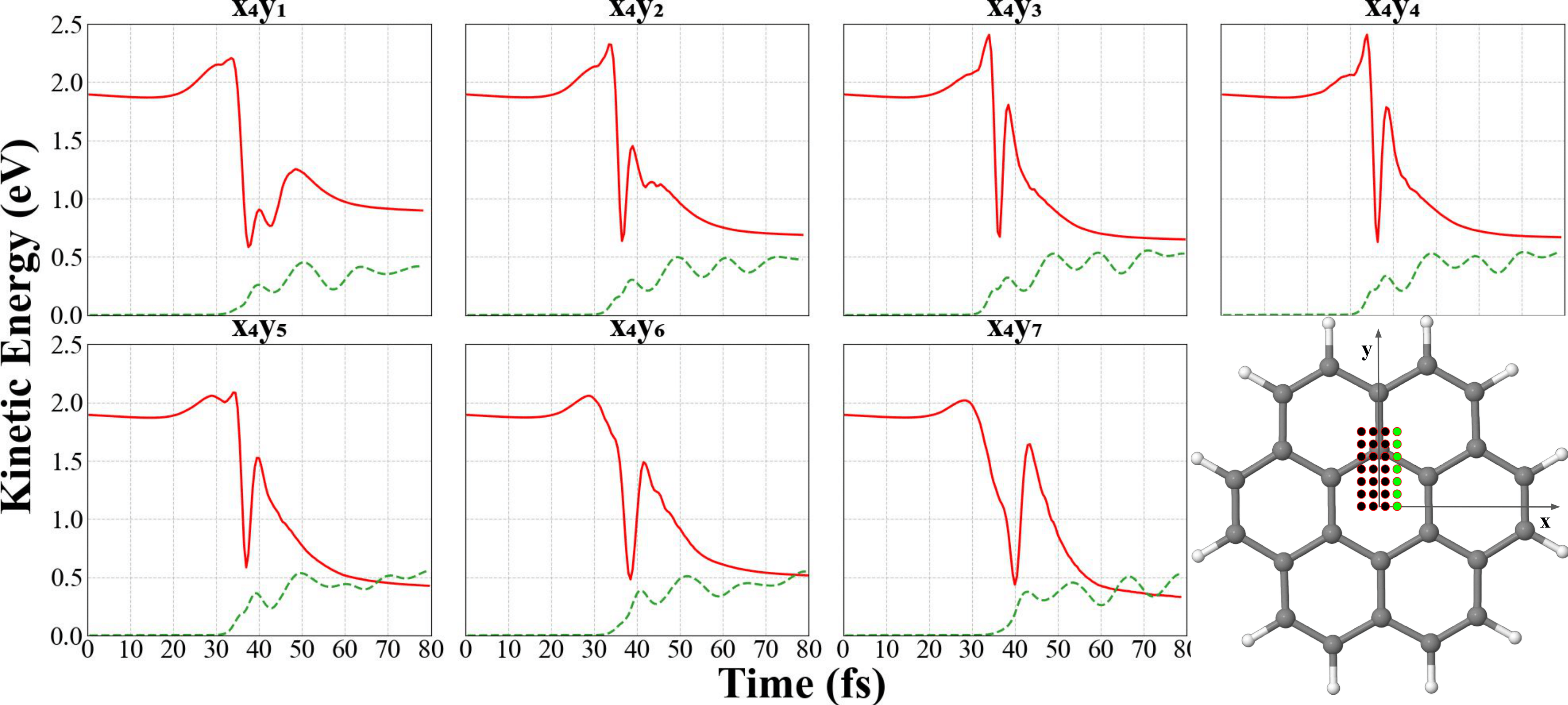}
\caption{Kinetic energy (in eV) of the H atom (red solid line) and the net kinetic energy of all atoms in the coronene sheet (green dashed line) as a function of time throughout the simulation. Each graph corresponds to a different simulation from the column of x\textsubscript{4} incident points, ranging from y\textsubscript{1} to y\textsubscript{7}.}\label{k-energy-x4}
\end{figure*}

Fig.~\ref{incident-point-diagram} illustrates the incident points where the hydrogen was aimed in each of the 28 simulations. The coronene sheet that the hydrogen was aimed at lies in the xy-plane. A $4 \times 7$ point grid of incident points was used for the x and y values, respectively, with a grid spacing of 0.3~\AA. In all simulations, the hydrogen projectile was launched at an angle of $27.4^\circ$ from the z-axis (orthogonal to the coronene plane) with a kinetic energy of 1.89~eV (velocity of 0.19~\AA/fs). These parameters for the initial kinetic energy and angle were motivated by a recent study~\cite{doi:10.1126/science.aaw6378}. The initial kinetic energy remained identical across all simulations, with components only in the x and z directions.

The initial positions of the H atoms were selected to align with the incident points illustrated in Fig.~\ref{incident-point-diagram}. The height above the coronene molecule was consistently maintained at 7~\AA\ for all simulations, ensuring uniformity in the initial z-coordinate. Additionally, the initial velocity and its angle were kept constant across all simulations. As a result, the only variable between simulations was the H atom's initial position in the xy-plane at $z = 5$~\AA, enabling the H atom to precisely target each of the incident points shown in Fig.~\ref{incident-point-diagram}.

Fig.~\ref{k-energy-x1} displays the kinetic energy (in eV) of the H atom throughout the simulations for all incident points along the x\textsubscript{1} column. In most cases, the H atom reaches the coronene surface at approximately 35~fs. As the H atom approaches the coronene (e.g., during 30~fs to 35~fs at incident point x\textsubscript{1}y\textsubscript{4}), its kinetic energy sharply decrease due to interactions with the coronene's potential energy barrier. In all simulations, the H atom failed to penetrate this barrier or form a bond with a C atom. Instead, the H atom experienced strong repulsion, leading to a sudden change in direction (e.g., from 35~fs to 40~fs at x\textsubscript{1}y\textsubscript{4}). After being repelled away from the coronene (beyond 40~fs), the H atom’s velocity gradually decreases, suggesting the presence of an attractive force exerted by the coronene. This deceleration stabilizes into a constant velocity once the H atom reaches a sufficient distance from the coronene, where it is no longer influenced by significant forces. 

The scattered H atom transferred a portion of its kinetic energy to the coronene structure, as evidenced in each kinetic energy graph where the total kinetic energy of all atoms in the coronene begins to increase upon interaction with the H atom at around 35~fs (Fig.~\ref{k-energy-x1}). This energy transfer is largely converted into vibrational energy within the coronene molecule, as indicated by the initial rise in net kinetic energy followed by oscillatory behavior observed after 40~fs.

Comparing the kinetic energy profiles of the H atom across all simulations in the x\textsubscript{1} column (Fig.~\ref{k-energy-x1}) provides important insights into the interaction dynamics. Notably, the interaction time between the H atom and the coronene sheet is significantly longer for incident points that are farther from the C atoms. For instance, at x\textsubscript{1}y\textsubscript{1}, the incident point furthest from the C atoms in the x\textsubscript{1} column, the interaction lasts approximately 20~fs. During this period, the H atom’s kinetic energy decreases before reaching the coronene (from 30~fs to 40~fs), followed by a prolonged escape phase after scattering (from 40~fs to 50~fs). In contrast, at points closer to the C atoms, such as x\textsubscript{1}y\textsubscript{7}, the interaction is much shorter. At x\textsubscript{1}y\textsubscript{7}, the velocity decreases around 35~fs, reaches a minimum at 38~fs, and then increases again until 40~fs. This comparison highlights that the interaction time at x\textsubscript{1}y\textsubscript{1} is roughly 20~fs\textemdash four times longer than the 5~fs interaction observed at x\textsubscript{1}y\textsubscript{7}, where the H atom interacts more briefly with the coronene.

H atoms aimed at incident points farther from C atoms experience longer interaction times because they are not immediately reflected or scattered upon approaching the graphene surface. This suggests that it is easier for the H atom to penetrate coronene's potential energy barrier at these incident points. Conversely, when the H atom is aimed directly at a C atom, the repulsion between the positively charged nuclei and their associated electron densities causes immediate reflection. This limits the time the H atom spends near the graphene plane, reducing its ability to overcome the potential energy barrier, rehybridize the C atom from sp\textsuperscript{2} to sp\textsuperscript{3}, and form a C-H bond.

This trend\textemdash incident points further from C atoms having a longer interaction period as opposed to immediate reflection\textemdash persists across all kinetic energy graphs, as shown in Fig.~\ref{k-energy-x1}, Fig.~\ref{k-energy-x2}, Fig.~\ref{k-energy-x3}, and Fig.~\ref{k-energy-x4}. In each of these figures, the y\textsubscript{1} incident points, which are furthest from any C atoms, exhibit significantly longer interaction periods, with changes in kinetic energy beginning around 30~fs and continuing until approximately 50~fs. In contrast, incident points near C atoms, such as x\textsubscript{2}y\textsubscript{6} and x\textsubscript{3}y\textsubscript{6}, demonstrate rapid, immediate repulsion and sharp kinetic energy changes.

The increased probability of C-H bond formation at these incident points is particularly evident at the x\textsubscript{4}y\textsubscript{1} incident point (Fig.~\ref{k-energy-x4}). In this graph, the H atom kinetic energy reaches a relative maximum at 40~fs. The kinetic energy decrease following the peak suggests the presence of additional attractive forces within the coronene that act on the H atom after it reaches the incident point. Although the H atom was ultimately scattered (as observed in all simulations across various incident points), the extended interaction period and the attractive forces from the electron densities of the C atoms allowed the H atom to remain near the coronene plane for a longer duration. This increased interaction time and attractive forces, shown in kinetic energy graphs for the x\textsubscript{1}y\textsubscript{1}, x\textsubscript{2}y\textsubscript{1}, x\textsubscript{3}y\textsubscript{1}, and x\textsubscript{4}y\textsubscript{1} (Fig.~\ref{k-energy-x1}-\ref{k-energy-x4}), enhances the probability of C-H bond formation.

\begin{table}[ht]
\begin{ruledtabular}
\begin{tabular}{lcccc}
    \textrm{} & \textrm{\boldmath$x_1$} & \textrm{\boldmath$x_2$} & \textrm{\boldmath$x_3$} & \textrm{\boldmath$x_4$} \\
    \colrule
    \textrm{\boldmath$y_1$} & 1.24 & 0.76 & 0.91 & 1.00 \\
    \textrm{\boldmath$y_2$} & 1.28 & 1.28 & 1.18 & 1.21 \\
    \textrm{\boldmath$y_3$} & 1.19 & 1.27 & 1.40 & 1.24 \\
    \textrm{\boldmath$y_4$} & 1.32 & 1.31 & 1.46 & 1.23 \\
    \textrm{\boldmath$y_5$} & 1.13 & 1.41 & 1.34 & 1.47 \\
    \textrm{\boldmath$y_6$} & 1.18 & 1.15 & 1.54 & 1.38 \\
    \textrm{\boldmath$y_7$} & 1.35 & 1.30 & 1.54 & 1.56 \\
\end{tabular}
\end{ruledtabular}
\caption{Energy differences (in eV) of the H atom between its initial kinetic energy (1.89~eV) and the kinetic energy after the collision, measured at each x\textsubscript{i}y\textsubscript{j} incident point.}
\label{hydrogen-energy-table}
\end{table}

\begin{table}[ht]
\begin{ruledtabular}
\begin{tabular}{lcccc}
    \textrm{} & \textrm{\boldmath$x_1$} & \textrm{\boldmath$x_2$} & \textrm{\boldmath$x_3$} & \textrm{\boldmath$x_4$} \\
    \colrule
    \textrm{\boldmath$y_1$} & 0.25 & 0.41 & 0.39 & 0.42 \\
    \textrm{\boldmath$y_2$} & 0.52 & 0.48 & 0.47 & 0.48 \\
    \textrm{\boldmath$y_3$} & 0.45 & 0.50 & 0.50 & 0.53 \\
    \textrm{\boldmath$y_4$} & 0.48 & 0.44 & 0.52 & 0.54 \\
    \textrm{\boldmath$y_5$} & 0.53 & 0.47 & 0.49 & 0.56 \\
    \textrm{\boldmath$y_6$} & 0.57 & 0.51 & 0.44 & 0.54 \\
    \textrm{\boldmath$y_7$} & 0.54 & 0.51 & 0.47 & 0.53 \\
\end{tabular}
\end{ruledtabular}
\caption{Vibrational energy differences (in eV) of the corenene between the ground state and after hydrogen collision at each x\textsubscript{i}y\textsubscript{j} incident point.}
\label{coronene-energy-table}
\end{table}

During the scattering interaction between the H atom and coronene, energy transfer occurs. The H atom loses kinetic energy, as shown in Fig.~\ref{k-energy-x1}--\ref{k-energy-x4}, by comparing its initial kinetic energy at 0~fs to its final kinetic energy at 80~fs. The exact values of kinetic energy loss for the H atom across all incident points are provided in Table~\ref{hydrogen-energy-table}. This lost energy is absorbed by the coronene and converted into vibrational motion of its atoms, as shown in Fig.~\ref{k-energy-x1}--\ref{k-energy-x4} as well. The energy transfer is further illustrated in Table~\ref{coronene-energy-table}, which displays the kinetic energy sum of all atoms (vibrational energy) in the coronene at 80~fs, highlighting the increase due to this interaction.

Based on Table~\ref{hydrogen-energy-table}, the calculated mean H atom kinetic energy loss is 1.27~eV. The kinetic energy loss exceeds 1~eV in all cases except for x\textsubscript{2}y\textsubscript{1}, x\textsubscript{3}y\textsubscript{1}, and x\textsubscript{4}y\textsubscript{1}, where the losses are 0.76~eV, 0.91~eV, and 1.00~eV, respectively. This lower kinetic energy loss at incident points further from C atoms is consistent with the previously discussed findings, where these points exhibit longer interaction times and are more likely to form a C-H bond. The smaller energy loss at these incident points suggests that the H atom transfers less energy to the coronene compared to points where the H atom is immediately reflected from the C atom it strikes, such as x\textsubscript{3}y\textsubscript{6}, where the H atom loses 1.54~eV (see Table~\ref{hydrogen-energy-table}). This observation aligns with the energy differences in coronene, as shown in Table~\ref{coronene-energy-table}, where the y\textsubscript{1} row of incident points (x\textsubscript{1}y\textsubscript{1}, x\textsubscript{2}y\textsubscript{1}, x\textsubscript{3}y\textsubscript{1}, and x\textsubscript{4}y\textsubscript{1}) exhibit the smallest energy differences. The average energy difference for all incident points, as detailed in Table~\ref{coronene-energy-table}, is 0.48~eV.

Notably, the energy differences for the H atom (Table~\ref{hydrogen-energy-table}) are much larger than those for coronene (Table~\ref{coronene-energy-table}). This discrepancy arises because not all of the kinetic energy of the H atom is transferred to coronene as vibrational energy\textemdash some of the energy is dissipated within the electron density. As shown in all the kinetic energy figures (Fig.~\ref{k-energy-x1}--\ref{k-energy-x4}), the vibrational energy of the coronene molecule increases upon interaction with the H atom. However, after the initial interaction, this energy oscillates as the molecules move, stretch, and repel each other. The increase in vibrational energy following the hydrogen scattering event results from the excitation of the coronene structure due to the energy transferred by the H atom.

\begin{figure}[ht!]
    \centering
    \includegraphics[width=0.95\columnwidth]{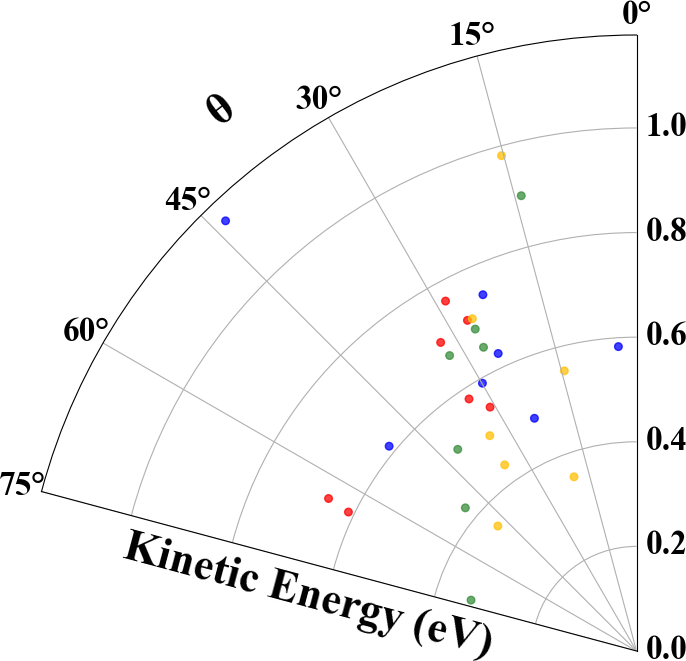}
    \caption{Angular distribution of scattered H atoms from the 28 simulations. The plot shows the final kinetic energy (at 80~fs) and the angle $\theta$, which is the angle between the velocity vector and the z-axis. The colors red, blue, yellow, and green correspond with simulations aimed at the x1, x2, x3, and x4 columns of incident points, respectively.}
    \label{fig:angular-distribution}
\end{figure}

The scattering angles and kinetic energies of the H atoms as they are deflected from the coronene surface are presented in Fig.~\ref{fig:angular-distribution}. This figure demonstrates the significant influence of the incident point on the scattering angle and final kinetic energy of the H atoms. Depending on the specific impact location, the hydrogen can be scattered nearly orthogonal to the coronene surface (e.g., x\textsubscript{2}y\textsubscript{4}, which reflects with an angle of $3.4^\circ$) or at a less orthogonal angle (e.g., x\textsubscript{4}y\textsubscript{7}, which scatters with an angle of $72.9^\circ$). The incident point x\textsubscript{2}y\textsubscript{1} results in the highest final kinetic energy (1.14~eV), while x\textsubscript{4}y\textsubscript{7} produces the lowest (0.33~eV). This wide variability in scattering angles and velocities underscores the critical role of the incident point in determining the H atom's trajectory and kinetic energy after interaction.

From the data, the mean and median final kinetic energy of the H atoms are 0.62~eV, and the mean deflection angle is $33.5^\circ$. These results indicate that, on average, the H atom loses 1.42~eV of kinetic energy after scattering, as corroborated by the mean of the values in Table~\ref{hydrogen-energy-table}. Furthermore, the data reveal that the hydrogen generally reflects at an angle closer to the coronene plane than its initial trajectory. In all simulations, the hydrogen approaches the graphene sheet at an angle of $27.4^\circ$ and, on average, reflects at an angle of $33.5^\circ$.

Interestingly, the scattering angles for the first row of incident points\textemdash previously identified as having a higher probability of adsorption\textemdash do not show a significant deviation from the overall average reflection angle. The reflection angles for incident points x\textsubscript{1}y\textsubscript{1}, x\textsubscript{2}y\textsubscript{1}, x\textsubscript{3}y\textsubscript{1}, and x\textsubscript{4}y\textsubscript{1} are $63.7^\circ$, $43.7^\circ$, $15.2^\circ$, and $14.2^\circ$, respectively. The average reflection angle for these points is $34.2^\circ$, closely aligning with the overall average reflection angle of $33.5^\circ$ across all incident points.

\subsection{Variable Kinetic Energy and Adsorption}

\begin{figure*}
\centering
\includegraphics[width=\textwidth]{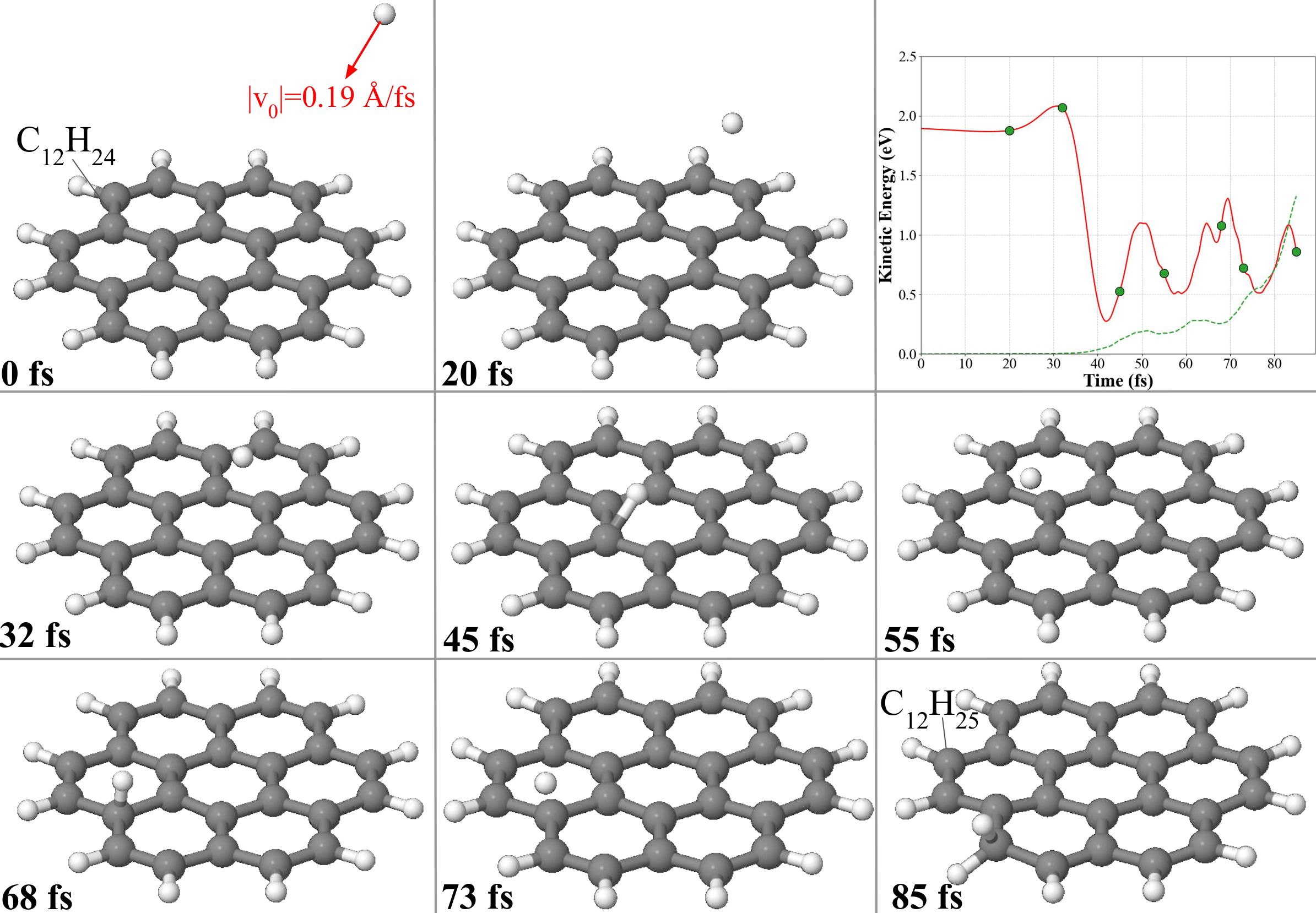}
\caption{Snapshots of a hydrogen projectile with an initial kinetic energy of 1.89 eV (velocity of 0.19 A/fs) being absorbed by a coronene molecule. The corresponding H atom kinetic energy (red solid line) is plotted over time, with green markers indicating the specific simulation times at which each snapshot was captured. The net kinetic energy of all the coronene atoms throughout the simulation is plotted as well (green dashed line)}
\label{fig:stick}
\end{figure*}

Fig.~\ref{fig:stick} presents snapshots from a simulation where the H atom was successfully adsorbed onto the graphene surface. In this simulation, the initial kinetic energy of the H atom was identical to that used in Fig.~\ref{k-energy-x1}-\ref{k-energy-x4} (1.89~eV), but the incident angle was adjusted to $35^\circ$ relative to the z-axis, making it more parallel to the xy-plane compared to the $27.4^\circ$ angle in the previous simulations. The H atom was targeted at an incident point of -0.82~\AA\ on the x-axis, chosen based on earlier findings that the y\textsubscript{1} row of incident points maximizes the probability of hydrogen adsorption. Additionally, the angle was altered to make the hydrogen approach closer to the C atoms, facilitating its ability to form a C-H bond after overcoming the coronene potential energy barrier.

As illustrated in Fig.~\ref{fig:stick}, the H atom initially breaks through the potential energy barrier of the coronene and forms a transient bond at 45~fs. At 68~fs, the hydrogen is displaced and temporarily bonds with another C atom. By the end of the simulation at 85~fs, the hydrogen forms a stable bond with a C atom at the edge of the coronene molecule. Although this final bonding is influenced by edge effects inherent to the coronene model, the H atom double-bounce trajectory and kinetic energy loss transfer following 45~fs suggests that, in a larger graphene system, a bond would likely form with a C atom in an interior benzene ring. 

Throughout this process, the H atom's kinetic energy is transferred to the coronene structure, as shown in the kinetic energy graph in Fig.~\ref{fig:stick}, by the increasing net kinetic energy of the coronene from 35~fs onward. This rise in energy corresponds to the vibrational excitation of the coronene atoms, which absorb the incoming hydrogen's kinetic energy. Unlike scattering interactions, where some of the kinetic energy remains with the hydrogen, in this adsorption scenario, the hydrogen's kinetic energy is entirely transferred to the coronene, initiating vibrational motion across the molecule.

\begin{table}[ht]
\begin{ruledtabular}
\begin{tabular}{lcccc}
    \textrm{Simulation} & \textrm{Initial KE} & \textrm{Result} & \textrm{Angle of} & \textrm{KE loss} \\
    \textrm{} & \textrm{(eV)} & \textrm{(A/S/T)} & \textrm{reflection ($^\circ$)} & \textrm{(eV)} \\
    \colrule
    \textrm{\textbf{1}} & 1.89 & A & ... & ... \\
    \textrm{\textbf{2}} & 2.07 & A & ... & ... \\
    \textrm{\textbf{3}} & 2.51 & A & ... & ... \\
    \textrm{\textbf{4}} & 3.50 & A & ... & ... \\
    \textrm{\textbf{5}} & 4.66 & S & 79.87 & 3.91 \\
    \textrm{\textbf{6}} & 6.35 & S & 57.34 & 3.45 \\
    \textrm{\textbf{7}} & 9.14 & T & -38.33 & 4.29 \\
\end{tabular}
\end{ruledtabular}
\caption{Outcomes of 7 simulations with varying initial kinetic energy of the H atom. The collision results in one of three events: adsorption (A), scattering (S), or transmission (T). For cases where the H atom does not form a bond with any C atom in the coronene (S and T), the angle of reflection and the kinetic energy loss are reported.}
\label{hydrogen-stick-velocity-test}
\end{table}

Seven simulations exploring the effect of varying the initial kinetic energy of the H atom are summarized in Table~\ref{hydrogen-stick-velocity-test}. In each simulation, the H atom was launched from the same initial position, with identical orientation and incident point on the coronene. Consistent with the setup depicted in Fig.~\ref{fig:stick}, the H atom was positioned 7~\AA\ above the coronene, aimed at x~=~-0.82~\AA, and incident at an angle of $35^\circ$ relative to the z-axis. Consequently, the first row in Table~\ref{hydrogen-stick-velocity-test} corresponds to the simulation presented in Fig.~\ref{fig:stick}.

Each simulation in Table~\ref{hydrogen-stick-velocity-test} produced a distinct outcome depending on the initial kinetic energy of the H atom. The four simulations with the initial H atom kinetic energy ranging from 1.89~eV to 3.50~eV resulted in adsorption (A), indicating that kinetic energies within this range at the given incident angle were sufficient to overcome the potential energy barrier of the graphene sheet and form a bond with one of the C atoms. In contrast, the two simulations with higher initial kinetic energies of 4.66~eV and 6.35~eV also overcame the potential barrier due to their increased energy but did so with excessive kinetic energy. This caused the H atom to approach too closely to the nuclear core of the C atom, resulting in scattering (S). The scattering behavior varied between these two cases, with the angles of reflection relative to the xy-plane provided in Table~\ref{hydrogen-stick-velocity-test}. Additionally, the amount of kinetic energy lost by the H atom during these simulations differed, with the higher-energy simulation (Simulation 6) losing less kinetic energy despite starting with a greater initial value.

Finally, in the simulation with the highest initial kinetic energy of 9.14~eV, the H atom had sufficient velocity to not only surpass the graphene's potential energy barrier but also transmit (T) through the sheet. In this case, 4.29~eV of the H atom's energy was absorbed by the graphene, and the H atom exited the sheet with a refracted angle of $-38.33^\circ$ relative to the xy-plane (where the negative angle indicates a velocity component below the xy-plane, corresponding to a negative z-component of velocity).

\section{Summary}
Time-dependent density functional theory (TDDFT) simulations were conducted to investigate the effects of varying incident points and velocities of H atom projectiles on a graphene-like structure. The findings reveal that the location where the H atom strikes the graphene significantly influences its kinetic energy, the vibrational motion of the graphene lattice, and the angles at which the H atoms are scattered.  

Clear trends emerged in the data based on the incident points. Incident points farther from C atoms result in longer interaction times between the H atom and the graphene, leading to smaller kinetic energy differences for both the H atom and the coronene structure after scattering. This extended interaction time and reduced energy transfer increases the likelihood of the H atom overcoming the graphene’s potential energy barrier, rehybridizing a C atom from sp\textsuperscript{2} to sp\textsuperscript{3}, and forming a covalent C-H bond.  

Building on these observations, a specific incident point and a velocity more parallel to the graphene plane were selected to achieve a successful simulation of H atom adsorption. Using this configuration, the effects of varying initial kinetic energies were tested. The results indicate that kinetic energies within the range of 1.89~eV to 3.50~eV enable adsorption. Higher kinetic energies (4.66~eV to 6.35~eV) allow the H atom to penetrate the potential energy barrier but result in strong repulsion from the positively charged carbon nuclear core, causing scattering. At much higher kinetic energies (9.14~eV and above), the H atom transmits through the graphene sheet, losing energy to the structure while refracting at a specific angle.  

These results highlight the critical role of initial parameters, such as incident points and kinetic energies, in determining the outcomes of the hydrogenation process of graphene. The findings underscore their impact on adsorption probability, scattering distributions, and energy transfer dynamics. Future work could focus on experimentally validating these results and further exploring the interplay of incident points and velocities.

\begin{acknowledgments} 
This work has been supported by the National Science Foundation (NSF)
under Grant No. DMR-2217759.

This work used ACES at TAMU through allocation PHYS240167 from the Advanced Cyberinfrastructure Coordination Ecosystem: Services \& Support (ACCESS) program, which is supported by National Science 
Foundation grants \#2138259, \#2138286, \#2138307, \#2137603, and \#2138296~\cite{aces}.
\end{acknowledgments}

%


\end{document}